\documentclass[varenna]{cimento}
\usepackage{epsf}
\title{Bose-Einstein Condensates in a 1D Optical Lattice}
\author{S. Burger, F.\,S. Cataliotti, F. Ferlaino, C. Fort,  P. Maddaloni, 
F. Minardi,  \atque M. Inguscio}
\institute{Laboratorio Europeo di Spettroscopia Nonlineare (LENS),\\
    Istituto Nazionale per la Fisica della Materia (INFM),\\
    Dipartimento di Fisica dell'\,Universit\`a di Firenze\\ 
    Largo~Enrico~Fermi,~2\\ 50\,125 Firenze, Italia  }
    
\begin{document}
\PACSes{\PACSit{03.75.Fi, 03.67.Lx, 32.80.Pj}{}}
%\PACSit{---.---}{\ldots}}
%03.75.Fi Phase coherent atomic ensembles; quantum condensation phenomena
%32.80.Pj Optical cooling of atoms; trapping
%32.80.Qk Coherent control of atomic interactions with photons
%67.57.De Superflow and hydrodynamics
%03.67.Lx Quantum computation

\maketitle
%\today  % <--------------------------------------------TAKE OUT AGAIN

\begin{abstract}
In this lecture we give an overview on current experiments
on Bose-Einstein condensation (BEC) in a one-dimensional (1D)
optical lattice. 
We introduce measurements of ground state, tunnelling and dynamical 
properties as well as investigations of atom optical applications 
and thermodynamical properties. 
Measurements of  the coherent atomic current in an array of
Josephson junctions and  the  critical
velocity for the breakdown of superfluid motion are discussed in 
detail.
\end{abstract}

\section{Introduction}

Bose-Einstein condensates (BECs) 
are quantum systems which can be easily 
manipulated and characterized due to their macroscopic nature~\cite{BECreview1}.
The employment of BECs in atomic physics has stimulated
a wealth of new experiments which is often compared to the rapid
development in optics and spectroscopy after the invention of the
laser~\cite{BECreview2}.

Atomic BECs confined to optical lattices
have been proposed for the realization of 
quantum computing schemes~\cite{Jaksch1998b,Brennen1999a}.
These approaches are of special interest because of the 
precise manipulation tools available in atomic physics.

Atoms confined in a periodic potential exhibit quantum effects known from
solid state physics, like Bloch oscillations and
Wannier-Stark ladders, which have been observed by exposing cold atoms to
the dipole potential of far detuned optical
lattices~\cite{Dahan1996a,Wilkinson1996a}.
The achievement of BEC has given the possibility to
explore also macroscopic quantum effects 
in this context.

In a first experiment with BECs loaded into a 1D optical lattice, 
quantum interference could be observed, leading to the formation of
the first ``mode-locked'' atom laser~\cite{Anderson1998a}.
The coherent nature of a BEC governs its dynamics
in optical lattices. 
From the well defined phase of the macroscopically occupied wavefunction
 describing 
the BEC  it follows that 
at low fluid velocities the BEC is performing a superfluid motion in the 
lattice~\cite{Burger2001a}.
In a regime of a greater potential height of the optical lattice,
the system is also ideally suited to study the Josephson effect:
At a potential depth of the lattice sites exceeding the thermal energy 
BECs  collectively
tunnel from one site to the next, at a rate 
which depends on the difference in phase between the 
sites. %while
Under the same conditions, 
thermal clouds of atoms are fixed to the wells of the 
optical lattice~\cite{Cataliotti2001a}.

Moreover, atoms confined to tightly confining potential wells
offer a controllable way to investigate 
ground state properties of the system~\cite{Pedri2001} and lower dimensional
physics, e.~g., the condensation process in in lower dimensions~\cite{Burger2001b}.

Recently,  also the squeezing of matter waves~\cite{Orzel2001a} 
and the decoherence of BECs in 2D optical lattices~\cite{Greiner2001} have
been investigated.
Experiments, in which optical lattices are applied to the BEC on much shorter 
time scales have investigated Bragg-diffraction as a tool for 
interferometry and 
spectroscopy~\cite{Kozuma1999a,Stenger1999b}, 
Bloch oscillations~\cite{Morsch2001}, 
and dynamical tunnelling~\cite{Hensinger2001a}.

Here we discuss recent experiments on 
macroscopic quantum effects of BEC dynamics in optical lattices.
The lecture is organized as follows:
In the chapter~\ref{setup_chapter} we briefly introduce the setup
for  the implementation of an optical lattice into
a BEC experiment.
Chapter~\ref{2D_chapter} concentrates on the Bose-Einstein phase
transition in the periodically modulated trap and 
discusses effects of two-dimensional (2D) physics.
Ground state properties of the coherent array of BECs in the optical 
lattice are reviewed in chapter~\ref{expansion_chapter}.
In chapter~\ref{SF_chapter} we discuss on the superfluid motion and
the density-dependent breakdown of superfluidity  of
a BEC  in an optical lattice.
Chapter~\ref{JJ_chapter} concentrates on the direct observation of
a coherent atomic current in an array of Josephson junctions.
Chapter~\ref{Concl_chapter} concludes the paper with an outlook
on future directions. 

%%%%%%%%%%%%%%%%%%%%%%%%%%%%%%%%%%%%%%%%%%%%%%%%%%%%%%%%%%%%%%%%%%%%%%%

\section{Experimental setup}
\label{setup_chapter}
Techniques for the achievement of BEC in dilute atomic gases
have been described in detail in Ref.~\cite{BECreview1}.
In the experiments discussed in the following chapters we create
BECs of $^{87}$Rb 
by the combination of laser cooling in 
a double magneto-optical trap system and evaporative cooling in a
static magnetic trap of the Ioffe-type~\cite{Fort2000a}.
The BECs  are produced
in the (F=1,m$_F$=\,$-1$) state, with
atom numbers of the order of $N\sim 10^6$. 
Due to the anisotropic magnetic trapping potential, 
the condensates are cigar-shaped with the long axis 
oriented horizontally; the 
typical dimensions (Thomas-Fermi radii)
are $R_x\sim 60\,\mu$m and $R_\perp\sim 6\,\mu$m.

\begin{figure}[t]
\epsfxsize=18pc 
\center{ \epsfbox{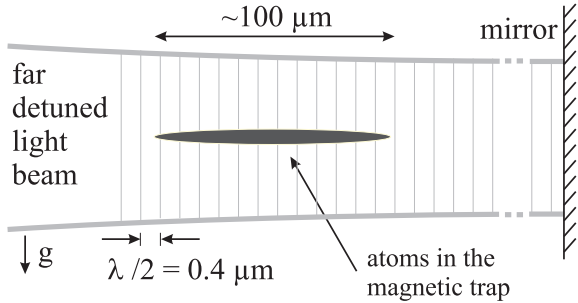}} 
\caption{Schematic set-up of the experiment.
\label{schema_v}}
\end{figure}

\begin{figure}[b]
\epsfxsize=18pc 
\center{ \epsfbox{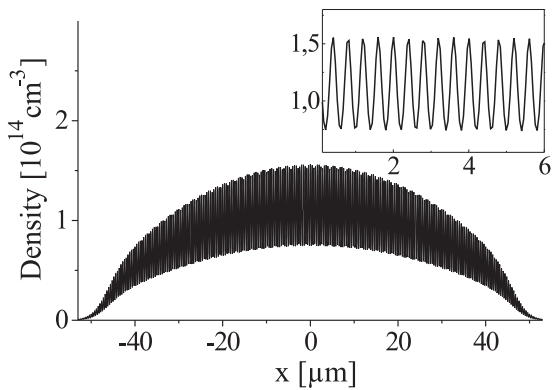}} 
\caption{ 
Linear density distribution of a BEC in the combined potential of the 
magnetic trap and optical lattice potential, obtained from a numerical simulation
with the parameters $N=3\times 10^5$ and $V_0=1.5\,E_R$.
\label{sim_gpe}}
\end{figure}

We create a 1D optical lattice by superimposing to the long axis
of the magnetic trap a far detuned,
retroreflected laser beam with wavelength $\lambda$
(see Fig.~\ref{schema_v}). 
The resulting
potential is given by the sum of the magnetic ($V_B$) and the optical
potential ($V_{opt}$):
\begin{equation}
V=V_B+V_{opt}= \frac{1}{2} m \left(\omega_x^2x^2+
\omega_{\perp}^2(y^2+z^2)\right) + V_0 \cos^2 kx \ ,
\label{optpot}
\end{equation}
where $m$ is  the atomic mass,
$\omega_x=2 \pi \times 9$~Hz and $\omega_{\perp}=2 \pi \times 90$~Hz
are the axial and radial frequencies of
the magnetic harmonic potential, 
and $k=2 \pi/ \lambda$ is the modulus of the  wavevector of the optical lattice. 
By varying the intensity of the
laser beam (detuned typically $\Delta= 150\,$GHz 
to the blue of the D$_1$ transition at
$\lambda=795$~nm) up to 14~mW/mm$^2$ we can vary the 
optical lattice potential height $V_0$ from 0 to $V_0\sim 5E_R$,
where $E_R$ is the recoil energy corresponding to the emission or absorption of 
one lattice photon, 
$E_R=\hbar^2 k^2/2m$.
To calibrate the optical potential we measure the
Rabi frequency of the Bragg transition between the
momentum states $-\hbar k$ and $+\hbar k$ induced by the standing
wave~\cite{Peik1997}.
Due to the large detuning of the optical lattice, 
spontaneous scattering can be neglected for
the experiments on BEC-dynamics which are performed typically on
a timescale of $\tau\sim 2\pi/\omega_x$;
nevertheless,  
spontaneous scattering leads to  a reduction of the
total atom number during the preparation of the BEC.

Bose-Einstein condensates in the combined magnetic trap and optical 
lattice are prepared by
superimposing the optical lattice to the trapping potential already 
during the last hundreds of ms of the RF-evaporation ramp. 
Figure~\ref{sim_gpe} shows the expected linear density
distribution of the ground state in a weakly binding lattice ($V_0=1.5\,E_R$), 
as obtained by numerical
propagation of the Gross-Pitaevskii equation in 
imaginary time. 
In the experiment, the
density  modulation on the length scale of $\lambda/2$ cannot be
directly resolved, due to the limited resolution 
of the imaging system.

%%%%%%%%%%%%%%%%%%%%%%%%%%%%%%%%%%%%%%%%%%%%%%%%%%%%%%%%%%%%%%%%%%%%%%%%
\section{Bose-Einstein phase transition lower dimensions}
\label{2D_chapter}

In this chapter we concentrate on the Bose-Einstein 
phase transition of the atomic gas 
in the combined potential of the
magnetic trap and the optical lattice
which allows to identify consequences of the 
reduced dimensionality of the system~\cite{Burger2001b}.

In atomic physics, important steps towards the realization of pure 2D
systems of neutral atoms have been made in different systems:
Significant fractions of atomic systems could be prepared in the 2D
potentials of optical lattices~\cite{Vuletic1998a,Bouchoule2001x} and
of an evanescent wave over a glass prism~\cite{Gauck1998a},
quasicondensates could be observed in 2D atomic hydrogen trapped
 on a surface covered with liquid $^4$He~\cite{Safonov1998a},
and 3D condensates of $^{23}$Na with low atom numbers 
could be transferred to the 2D
regime by an adiabatic deformation of the trapping 
potential~\cite{Gorlitz2001x}. 

By using Bose-Einstein condensates (BECs) trapped in optical lattices it has
become possible to overcome major limitations
of previous experiments:
First, an optical lattice can confine a large array of
2D systems, which allows measurements with a much higher number
of involved atoms with respect to a single confining potential. 
Second, the macroscopic population of a single quantum state (BEC)
naturally transfers the whole system to a pure occupation of the 
2D systems, which could so far not be realized with thermal
atomic clouds.

\begin{figure}[t]
\epsfxsize=12pc 
\center{\epsfbox{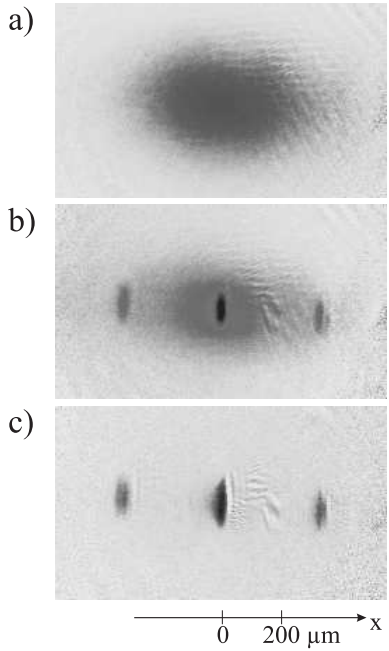} }
\caption{
Absorption images of a thermal cloud (a),  
a mixed cloud (b), and a 
pure BEC (c), expanded for 26.5\,ms from the combined magnetic
trap and optical lattice ($V_0\sim 4 E_R$).
The corresponding temperatures and atom numbers are 
$T\approx 210\,$nK and $N\approx 4\times 10^5$ (a), 
$T\approx 110\,$nK and $N\approx 1.5\times 10^5$ (b), 
$T < 50\,$nK and $N\approx 1.5\times 10^4$ (c).}
%Data from 19.3.2001:nr.31,33,35
\label{absbild}
\end{figure}

The dimensionality of a gas of weakly interacting bosons has important
consequences for its thermodynamical properties.
While in 3D the gas undergoes the phase transition to BEC 
even in free space, in 2D systems BEC at finite temperatures can only exist
in a confining potential~\cite{Petrov2000a}.
For the ideal gas in a 2D harmonic trap with the fundamental frequency
$\omega$ the analytical solutions for
the condensation temperature, $T_c$, and the dependence of the condensate
fraction, $N_0/N$ (number of particles in the ground state, $N_0$, and total
particle number, $N$) are given by~\cite{Bagnato1991a}:
\begin{equation}
k_B T_c = \hbar \omega \left( \frac{N}{\zeta(2)} \right)^{1/2}\ ,\hspace{0.8cm}
\frac{N_0}{N}=1-\left( \frac{T}{T_c} \right)^2\ ,
\label{N_N_eqn}
\label{T_c_eqn}
\end{equation}

where $\zeta(s)$ is the {\it zeta}-function, defined as
$\zeta(s)=\sum_{n=1}^{\infty}1/n^s$.
In the 3D case these dependencies are:
\begin{equation}
k_B T_c = \hbar \omega \left( \frac{N}{\zeta(3)} \right)^{1/3}\ ,
\hspace{0.8cm}\frac{N_0}{N}=1-\left( \frac{T}{T_c} \right)^3\ .
\label{N_N_eqn3D}
\end{equation}

In our experiment, the magnetic trapping potential is a 3D potential which confines the 
atoms to an overall cigar-shaped distribution,  
while in the 1D optical lattice the atoms are confined to 
2D planes.
Therefore, by increasing the strength of the optical lattice
superposed to a 3D potential it is possible to follow the transition from
a 3D BEC to an array of 2D degenerate atomic clouds 
confined radially by the magnetic
potential and assorted in the axial direction like disks in a shelf.

For reaching the quasi 2D regime, the motion of the particles 
has to be effectively ``frozen'' in the direction of the optical lattice
beam~\cite{Petrov2000a}, i.e., the fundamental frequency in a single 
lattice site, $\omega_l$ has to fulfill $\hbar \omega_l\gg k_B T$.
For our experimental parameters of $T<200\,$nK and 
$\omega_l\approx 2\pi 14\,$kHz (for $V_0\approx 4\,E_{rec}$) this relation
is well satisfied. 
Nevertheless, due to the small width of the barriers 
atoms can tunnel between the lattice sites.
The low energy of thermal atoms allows them to tunnel only over a few
sites during the duration of the experiment. 
Therefore we expect only minor changes of the thermodynamic properties due 
to such processes.
In contrast, tunnelling of ground state atoms is greatly enhanced because 
the ground state is macroscopically occupied (see chapter~\ref{JJ_chapter}).
As a result, the BECs at the optical lattice sites form a phase
coherent ensemble giving rise to the interference
pattern in the expansion (see chapter~\ref{expansion_chapter}).

\begin{figure}[t]
\epsfxsize=18pc 
\center{ \epsfbox{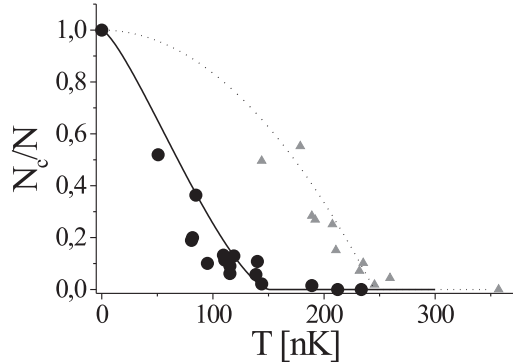} }
\caption{
Ground state occupation vs. temperature in the combined trap with
$V_0\approx 4\,E_{rec}$ (circles), and in the 3D purely magnetic trap (triangles).
The solid (resp. dotted) line gives the expected dependence for the combined
(resp. purely magnetic) trap, according to Eqn.~\ref{N_N_eqn} 
(resp. Eqn.~\ref{N_N_eqn3D}).}
%Data from 19.3.2001
\label{Nc_N_fig}
\end{figure}

In order to measure the effects of dimensionality 
on $T_c$ and on $N_0/N (T)$ we have
adjusted the final temperature of the atomic clouds  
by evaporative cooling
and recorded the atomic density distributions at different temperatures. 
Figure~\ref{absbild} shows  absorption images of such ensembles, 
expanded from a combined trap with a lattice-potential height
of $V_0\approx 4\,E_{rec}$.
The interference pattern of the expanding array
of BECs 
appears in three
spatially separated peaks (in Fig.~\ref{absbild}\,b,c)~\cite{Pedri2001}.

By integrating over the atomic density distribution we 
obtain the number of atoms in the ground state, $N_0$, 
and in the thermal cloud, $N_{th}=N-N_0$.
Figure~\ref{Nc_N_fig} shows  the ground state
fraction $N_0/N$, in dependence of the temperature of the ensemble.
In the case of the 3D potential of the pure magnetic trap 
(triangles in Fig.~\ref{Nc_N_fig}) 
this ratio reproduces the shape expected from Eqn.~\ref{N_N_eqn3D}
(dotted line, using a linear fit to the measured function $N(T)$).
The shape of the curve for ensembles produced in the
combined trap is  much smoother around the (lower) transition temperature and mixed
clouds with a relatively small condensate fraction exist in a broad
temperature range well below $T_c$.

This behaviour can be qualitatively understood with
a simplifying model~\cite{Burger2001b} assuming the subsequent
formation of 2D BECs at the different lattice sites: 
Due to the magnetic trapping potential the central lattice wells are
populated with a higher number of atoms, 
which -- according to Eqn.~\ref{T_c_eqn} --
leads to a higher critical temperature for the central clouds than for the
clouds in the wings of the overall density distribution.
BECs form first in the central 2D disks,
lowering the temperature leads to BEC formation at more and more lattice 
sites. 
The solid line in Fig.~\ref{Nc_N_fig} shows the ground state occupation according 
to this model; the curve agrees well with the experimental data points. 
A more sophisticated theoretical modelling of the problem, 
including interactions and the effect of tunnelling on the
thermodynamic properties, remains to be developed.

%%%%%%%%%%%%%%%%%%%%%%%%%%%%%%%%%%%%%%%%%%%%%%%%%%%%%%%%%%%%%%%%%%%%%%%%%%%%%

\section{Expansion of a BEC from the combined trap}
\label{expansion_chapter}

\begin{figure}[b]
\epsfxsize=15pc 
\center{ \epsfbox{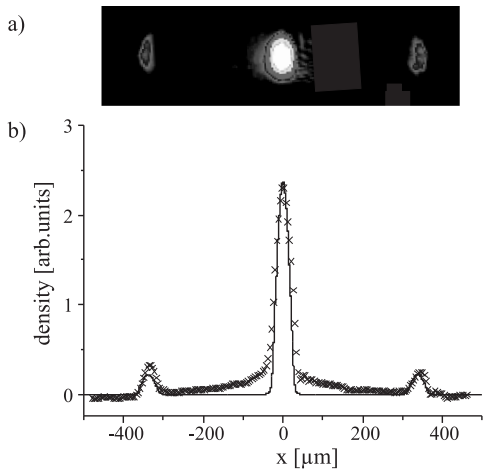} }
\caption{
a) Absorption image of the expanded array of condensates, showing three 
interference peaks, $n=-1,0,1$.
b) Density profile obtained from the absorption image (crosses)
and calculated density profile for the experimental parameters
$V_0=5E_R$ and $t_{exp}=29.5\,$ms~\cite{Pedri2001}.
The wings of the central peak in the experimental density profile result
from a small thermal component. 
}
\label{foto}
\end{figure}

In this chapter we introduce  ground state
properties of  the  fully coherent array of
condensates in the optical lattice. 
To this aim we explore the interference
pattern in the expanded cloud, reflecting the initial geometry of
the sample.
The time
evolution of the interference peaks, their relative population as well as the
radial size of the expanding cloud are compared to 
theoretical results developed in~\cite{Pedri2001}. 

\begin{figure}[t]
\epsfxsize=32pc 
\center{ \epsfbox{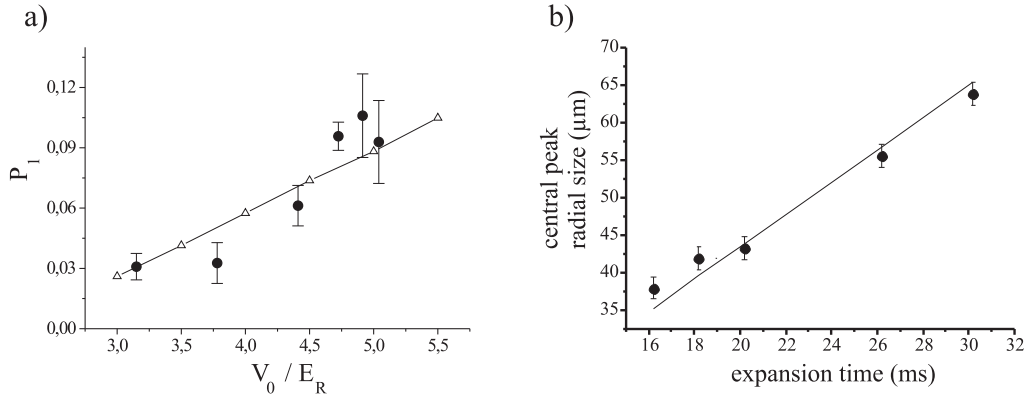} }
\caption{
a) Experimental (circles) and theoretical values (triangles)
of the relative population of the $n=1$\,peak with respect
to the central one ($n=0$) as a function of the lattice
potential $V_0$ in units of the recoil energy $E_R$.
b) Radial size of the central peak as a function of the 
expansion time. Experimental data points are compared to 
the expected asymptotic law $R_\perp(t)=R_\perp\omega_\perp t$.
}
%\label{efficienza_fig}
\label{exp_fig}
\end{figure}

In analogy of multiple order interference fringes in light
diffraction from a grating, the expansion of an array 
of coherent BECs leads 
for long expansion times (i.e., in the far field) to
distinct peaks in position space, reflecting the momentum
distribution.
The analogy is best
understood considering a periodic and coherent array of 
condensates aligned along the $x$-axis. 
The momentum distribution of the
whole system  is affected
in a profound way by the lattice structure and exhibits
distinctive interference phenomena (see~\cite{Pedri2001}). 
The momentum distribution is characterized by sharp peaks
at the values $p_x=2n \hbar k$ with $n$ integer (positive or
negative) whose weight is modulated by a function $n_0(p_x)$.

In Fig.~\ref{foto}a we show a typical image of the cloud taken at
t$_{exp}=29.5\,$ms with a total number of atoms $N=20000$ 
and a potential height of $V_0=5E_R$. 
The structure of the observed density profiles is well reproduced
by the free expansion of the ideal gas assuming a periodic ground state wave
function $\Psi$ describing the array of BECs~\cite{Pedri2001}. 
A predicted result for the density distribution
$n(x)=|\Psi(x) |^2$ evaluated for the experimental parameters  is
shown in Fig.~\ref{foto}b (continuous line).

From the experimental images  
we  determine the relative population of
the lateral peak with respect to the central one. 
The results for the relative population of the first
lateral peak as a function of the optical lattice potential $V_0$ are shown
in Fig.~\ref{exp_fig}\,a.
This figure shows also the
corresponding results of 
the 1D theoretical model developed in~\cite{Pedri2001}. 

More general,  it can be shown
that the relative population of the $n \neq 0$ peaks with
respect to the central one ($n=0$) obeys the simple law
\begin{equation}
P_n=\exp[-4 \pi^2 n^2 \sigma^2/d^2]
\label{pop}
\end{equation}
where $\sigma$ characterizes the width of a single condensate in a 
lattice site and $d$ is the periodicity of the lattice, $d=\lambda/2$.
Equation~\ref{pop} holds also in the presence of a smooth modulation of the atomic
occupation number $N_k$ in each well.  
Result (\ref{pop}) shows
that, if $\sigma$ is much smaller than $d$ the intensity of the
lateral peaks will be high, with a consequent important layered
structure in the density distribution of the expanding cloud. 
The value of $\sigma$ is determined, in first approximation,
by the optical confinement. 
It can be obtained 
numerically giving
$\sigma/d = 0.30 $, $ 0.27 $, and $0.25$ for $V_0=3$, $4$ and $5\,E_R$
respectively~\cite{Pedri2001}.

A 3D model  permits also to  explain the behaviour of
the radial expansion of the gas. In the presence of the density
oscillations produced by the optical lattice the problem  is not
trivial and should be solved numerically by integrating the GP
equation. 
However, after the lateral peaks are formed, the
density of the central peak expands smoothly according to the
asymptotic law $R_{\perp}(t) = R_{\perp}(0)\omega_{\perp}t_{exp}$,
holding for a cigar configuration in the absence of the optical
lattice \cite{castin}. 
As can be seen from Fig.~\ref{exp_fig}\,b the experimental values 
of the radial size of 
the peak after varied expansion time  compare well to this law~\cite{Pedri2001}.

Concluding, we  remark that  the well understood behaviour of the expanding 
cloud can be used as a  tool to investigate the coherence
properties of the BEC. 
Deviations from the shape of the interference pattern and from the
population of the interference peaks can be interpreted to result, e.g., 
from decoherence effects. 
In recent experiments, the interference peaks have been used to 
prove the well-defined phase relation in an array of Josephson 
junctions~\cite{Cataliotti2001a}.
The disappearence of the interference
peaks has been used to study squeezing of matter waves~\cite{Orzel2001a}.
Further studies include the possible effects of thermal decoherence
 in the presence of tighter optical traps~\cite{pitaevskii01,tognetti}.

%%%%%%%%%%%%%%%%%%%%%%%%%%%%%%%%%%%%%%%%%%%%%%%%%%%%%%%%%%%%%%%%%%%%%%%%%%%%%%%

\section{Superfluid dynamics} 
\label{SF_chapter}

Superfluidity of BECs is a direct consequence of their coherent 
nature~\cite{Dalfovo1999a}.
It is manifested in the appearance of vortices~\cite{Matthews1999b,Madison2000a}
and scissors modes~\cite{Marago2000a} as well
as in a critical velocity for the onset of
dissipative processes~\cite{Raman1999a}.
A far detuned optical lattice 
at a low potential height, $V_0<2\,E_R$, is well suited to study in detail 
the critical velocity because it acts like a medium with a 
microscopic roughness on the BEC moving through it, being velocity-dependent
compressed and decompressed as it propagates.

In order to investigate the dynamics in the combined trap we translate
the magnetic trapping potential in the $x$-direction by a
variable distance $\Delta x$ ranging up to $300\,\mu$m
in a time  $t\ll 2\pi/\omega_x$. 
Therefore,
the BEC finds itself out of equilibrium and is forced 
into motion by a potential gradient. 
After an evolution time $t_{ev}$
in the displaced trap, both the magnetic trapping and
the optical lattice are switched off and 
after a free expansion of 26.5\,ms the cloud is imaged, 
giving information on the momentum distribution of the system.

In the magnetic trap
the center-of-mass motion of the BEC
in the displaced trap is
an undamped oscillation with frequency $\omega_x=2\pi\times
8.7\,$Hz and amplitude $\Delta x$. 
In the combined trap formed by the magnetic 
and the (weakly confining, $V_0\sim 1.5\,E_R$) 
optical lattice potentials we observe dynamics in different regimes:\\
For small displacements, $\Delta x< 50\,\mu$m, the
dynamics of the BEC resembles the ``free oscillation'' at the same
amplitude but with a significant shift in frequency
which can be explained in terms of an effective atomic mass~\cite{Burger2001a}.
By varying the potential height $V_0$ we are able to tune this
effective mass.
The undamped  dynamics without dissipative processes 
 in the small-amplitude regime is a manifestation of
superfluid behavior of the BEC. 
When we further increase the initial 
displacement $\Delta x$ and hence the velocity of the
BEC, it enters a regime of dissipative dynamics. 
We observe a damped oscillation in the trap and dissipative
processes heating the cloud.

\begin{figure}[t]
\epsfxsize=32pc 
\center{ \epsfbox{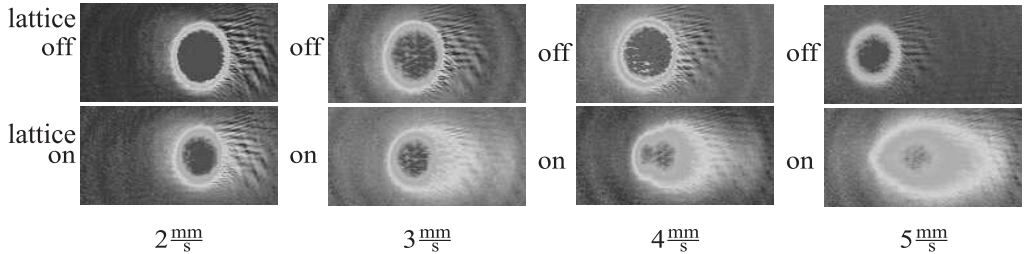} }
\caption{
False color representation of absorption images of atomic clouds
after evolution with different maximum velocities (indicated) in
the magnetic trap (``lattice off'') and in the combined trap (``lattice on''). 
\label{fig_sf1}}
\end{figure}

The critical velocity in a superfluid is proportional to 
the local speed of sound, $c_s$,  which depends on the
density $n$, $c_s(r)=\sqrt{n(r)/m\,(\delta\mu/\delta n) }$,
with  the chemical potential $\mu$. 
Therefore superfluidity breaks down first in the wings of the BEC
where the density is lowest.

\begin{figure}[t]
\epsfxsize=14pc 
\center{ \epsfbox{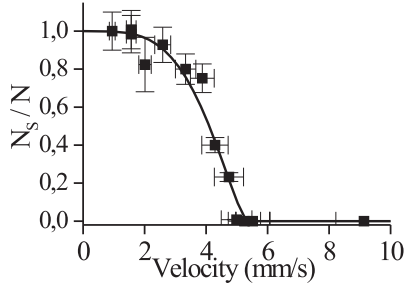} }
\caption{
Fraction of atoms in the superfluid component vs. maximum propagation 
velocity.
\label{fig_sf2}}
\end{figure}

In order to measure the velocity- and density-dependent 
onset of dissipation 
and thereby the spectrum of critical velocities in the BEC,
we have varied the displacement $\Delta x$ and
recorded atomic distributions after a fixed evolution time
$t_{ev}=40\,$ms. For low velocities, $v<2\,$mm/s,  
the sample follows the position of a freely moving BEC
(``lattice off'' in Fig.~\ref{fig_sf1});
no thermal component appears.

Upon increasing the velocity of the BEC, we observe a retardation of a part of
the cloud, leading to a well detectable separation 
from the superfluid component after free
evolution (see Fig.~\ref{fig_sf1}). 
For velocities $v\sim 4\,$mm/s we observe that
only the central part of the fluid is moving without
retardation; 
For velocities $v\sim 4\,$mm/s we observe that
only a  part of the fluid is moving without
retardation; the central position of this part 
is the same as the central position of the ``freely oscillating'' 
BEC.
For even higher velocities 
all of the atoms are retarded and form 
a heated cloud with a Gaussian density distribution.
The spatial separation 
from the thermal component allows a clear demonstration 
of the superfluid properties of inhomogeneous Bose-Einstein
condensates.

Figure~\ref{fig_sf2} shows the ratio of atom number in the non-retarded
component (parabolic density-profile, ``superfluid component''), $N_s$, and
the total atom number, $N$, in dependence of the maximum velocity
attained during the evolution in the optical lattice~\cite{footnote1}. 
The envelope function of the density
distribution of the BEC is an inverted parabola in 3D (see
Fig.~\ref{sim_gpe} and hence,  by 
integration over the high-density region,  we get an equation for
the relative number of atoms in the superfluid part of the BEC
for a given velocity~$v$,
$N_s(v)/N=\left[5/2\times(1-v^2/v_{max}^2)^{3/2}-3/2\times
(1-v^2/v_{max}^2)^{5/2}\right]$,
where $v_{max}$ is the critical velocity at maximum density.
This expression implies that about 90\% of the atomic probability
density is localized in a region which remains superfluid up to
velocities $v \simeq v_{max}/2$.
The line in Fig.~\ref{fig_sf2} shows that the above expression for 
$N_s(v)/N$ gives a very good account of the data, the fitted 
value of the maximum velocity being $v_{max}= (5.3 \pm 0.5)\,$mm/s. 

In the regime of a weak optical lattice potential, the lattice acts with
a rather small perturbation on the BEC, the atoms 
are completely delocalized within the macroscopic BEC wavefunction. 
However, increasing the 
potential we reach a regime where the atoms are confined to single
lattice sites. 
Here, the ensemble is better described 
in terms of an array of discrete BECs, connected to each other by 
tunnelling.
This regime is adressed in the following chapter.

%%%%%%%%%%%%%%%%%%%%%%%%%%%%%%%%%%%%%%%%%%%%%%%%%%%%%%%%%%%%%%%%%%%%%

\section{Observation of a Josephson current in an array of coupled BECs} 
\label{JJ_chapter}

Two macroscopic quantum systems which are coupled by a weak
link produce the flow of a supercurrent $I$ between them, driven by
their relative phase $\Delta \phi$,
\begin{equation}
\label{current-phase}
I=I_c \sin{\Delta \phi}\, ,
\end{equation}
where $I_c$ is the critical Josephson  current~\cite{Barone1982,Smerzi1997}.
The relative phase
evolves in time proportionally to the difference in chemical potential
between the two quantum fluids.

The first experimental
evidence of a current-phase relation was already obtained in
superconducting systems soon after
Josephson's proposal~\cite{Josephson1962}.
Also, phase-coherent tunnelling of BEC atoms from an optical lattice to the 
continuum, driven by gravity, has been observed~\cite{Anderson1998a}.
We realize a one-dimensional array
of bosonic Josephson junctions (JJs) by 
preparing an array of BECs in the sites of 
the optical lattice with an interwell barrier energy $V_0$ which is
high compared to the chemical potential of the BECs~\cite{Cataliotti2001a}.
Every two condensates in neighbouring wells 
overlap slightly with each other due to a finite tunnelling probability, 
and therefore constitute a
JJ, with the possibility to adjust the critical current $I_c$ by tuning the laser
intensity.
By driving the system with
the external harmonic potential, we 
investigate the current-phase dynamics and measure 
the critical Josephson current as a function of the
interwell potential~$V_0$.

\begin{figure}[t]
\epsfxsize=32pc 
\center{ \epsfbox{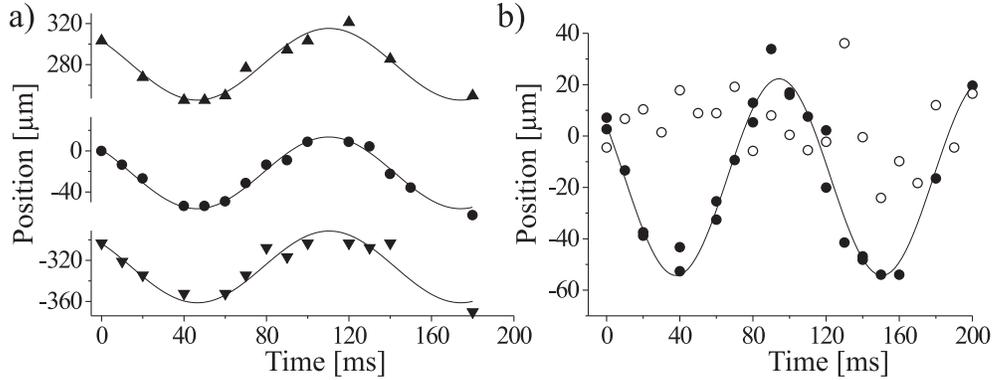} }
\caption{
a)  
The three peaks of the interference pattern
of an array BECs, expanded for 28\,ms after the
propagating in the optical lattice.
b) 
Center-of-mass positions of thermal clouds after 
expansion from the magnetic trap (filled circles)
and from the combined magnetic trap and optical lattice
(open circles)
as a function of evolution time in the displaced respective trap.
\label{fig_jj_osc}}
\end{figure}

\begin{figure}[b]
\epsfxsize=32pc 
\center{ \epsfbox{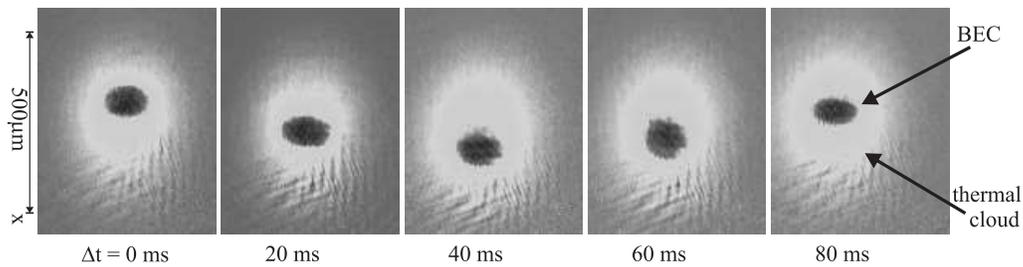} }
\caption{
False color representation of absorption images of mixed atomic clouds
in the displaced magnetic trap with superimposed optical lattice.
In the time evolution $\Delta$t  the ground state is moving relative 
to the thermal cloud.}
\label{fig_bec_th_jj}
\end{figure}

In its ground state the system  
consists of spatially separated condensates; tunneling between
adjacent wells leads to a constant phase over the whole array.
Therefore, the condensates show an interference pattern after
an expansion from the combined trap  (see Fig.~\ref{foto}).   
%This provides us with information about the
%relative phase of the different condensates.

\begin{figure}[t]
\epsfxsize=18pc 
\center{ \epsfbox{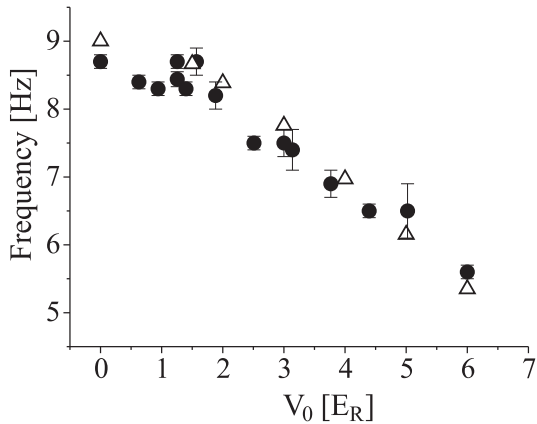} }
\caption{
Frequency of the oscillation of a coherent ensemble
 in the JJ-array in dependence of the interwell potential height.
Filled circles: experimental data, open triangles: results from 
a numerical simulation of the Gross-Pitaevskii 
equation~\cite{Trombettoni_phd}.
\label{fig_jj_freq}}
\end{figure}

To observe a Josephson current in the array we non-adiabatically  
displace the magnetic trap along the lattice axis by a 
distance of $\sim 30\, \mu m$.
The potential energy the atoms gain in this process is much smaller
than the interwell potential barrier, but the relative phases of 
the BECs in the different wells are driven by this process.
According to equation~\ref{current-phase} we expect a Josephson current.
A collective motion can be
established only with a well defined phase relation between
the condensates. 
This locking of the relative phases
shows up in the expanded cloud interferogram.
The expected Josephson current is observed as the collective oscillation
of the atomic ensemble. 
In Fig.~\ref{fig_jj_osc}a the positions of 
the three peaks in the interferogram are plotted 
as a function of time spent in the
combined trap after the displacement of the magnetic trap.
The motion performed by the center of mass of the ensemble is an
undamped oscillation at a frequency~$\omega<\omega_x$.

The coherent nature of the
oscillation is also proven by repeating the same experiment with a thermal
cloud. 
In this case -- although atoms can  individually tunnel
through the barriers -- no macroscopic phase is present in the cloud and
no motion of the center of mass is observed.  
The center-of-mass positions of thermal clouds in the optical lattice
are shown in Fig.~\ref{fig_jj_osc}b,
together with the oscillation of thermal clouds in absence
of the optical potential.  
As can clearly be seen, the movement of  thermal clouds
is strongly suppressed in presence of the optical
lattice.
We have also subjected mixed clouds to the displaced potential, 
where only the condensate fraction starts to
oscillate while the thermal component remains static; the
interaction of the two eventually leads to a damping of the
condensate motion and a heating of the system.
Figure~\ref{fig_bec_th_jj} shows absorption images of mixed clouds, 
where  a 
BEC moves relative to the thermal cloud.

As can be derived from the phase-current relation of the 
JJ array~\cite{Cataliotti2001a} 
the critical Josephson current is related to the small amplitude
oscillation frequency $\omega$ of the JJ array by the simple relation
$
I_c = \frac{4 \hbar}{m \lambda^2} 
\left( \frac{\omega}{\omega_x} \right)^2$.
Figure~\ref{fig_jj_freq} shows experimental values of the
oscillation frequency $\omega$ together with results from  
numerical solutions of the Gross-Pitaevskii equation. 
The possibility to precisely adjust the critical Josephson  current presents
a major advantage of Josephson junctions in Bose-Einstein condensates, 
where due to the elaborate manipulation tools of atomic physics a variety
of parameters can be tuned, compared to systems realized in solid-state physics.

\section{Conclusions} 
\label{Concl_chapter}

In this lecture we have discussed
thermodynamical and ground state properties of a
dilute gas in the potential of an optical lattice.
We have also investigated macroscopic quantum effects
in the dynamics of the system, such as 
superfluid motion, a density-dependent critical velocity for
the onset of dissipation, and -- in a regime of higher lattice potentials --
an oscillating Josephson current.

Future directions of this work include the study of 
BECs in 2D and 3D optical lattices and in  superlattices, 
which offers the possibility of a better control of the 
atom number per lattice site and of the
manipulation of single lattice sites, 
and possible applications of quantum computing schemes.
We also plan to further study BEC properties in these
systems, such as the dynamical behaviour of BECs in 
lower dimensions or the formation of bright atomic solitons
in optical lattices~\cite{Zobay1999a}.

\acknowledgments
Our understanding of the various experiments performed at LENS 
benefitted very much from the theoretical support in  collaborations with
M.\,L.~Chiofalo, P.~Pedri, L.~Pita\-ev\-skii,
 A.~Smerzi, S.~Strin\-gari, M.~Tosi, and A.~Trombettoni. 
We acknowledge stimulating discussions with
M.~Artoni, G.~Ferrari, and G.\,V.~Shlyapnikov.
We also acknowledge support by  the EU under contracts
HPRI-CT 1999-00111 \& HPRN-CT-2000-00125,
by MURST through  the PRIN\,1999 \& PRIN\,2000  Initiatives,
and by INFM under the contract {\it ``Photon Matter''}.


\begin{thebibliography}{0}
\bibitem{footnote1} 
Due to spontaneous scattering during the preparation
of the BEC in the combined trap the total atom number is reduced
by a factor $\sim 0.7$.
\end{thebibliography}
\end{document}